\documentclass[conference]{IEEEtran}
\IEEEoverridecommandlockouts
\usepackage{url}
\usepackage{graphicx}
\usepackage{import}
\usepackage{xcolor}
\usepackage{listings}

\begin{document}

\title{The Bipartible Architecture Pattern: Transparently Directing HTTP Requests over InfiniBand}

\author{
\IEEEauthorblockN{Richard Henwood}
\IEEEauthorblockA{\textit{Dell Technologies} \\
1 Dell Way, Round Rock, TX, USA \\
Richard.Henwood@dell.com}
\and
\IEEEauthorblockN{John Lockman}
\IEEEauthorblockA{\textit{Dell Technologies} \\
1 Dell Way, Round Rock, TX, USA \\
John.Lockman@dell.com}
\and
\IEEEauthorblockN{Luke Wilson}
\IEEEauthorblockA{\textit{Dell Technologies} \\
1 Dell Way, Round Rock, TX, USA \\
Luke.Wilson@dell.com}
}

\maketitle

\begin{abstract}

Large-scale compute (HPC) systems with coherent-memory workloads often deploy
InfiniBand (IB) networks for high-bandwidth and low-latency inter-node and
storage communication. A separate TCP network is typically used for
control-plane and general-purpose communication. The last decade has seen the
rise of cloud-native software patterns. These patterns are characterized by
container-based software distribution and inter-node and storage communication
over TCP/IP. In an unoptimized HPC environment, a cloud-native workload will
typically use HTTP for some or all communication. 


This paper introduces the Bipartible architecture pattern to optimize
cloud-native workload traffic in HPC environments. This paper focuses on
optimizing HTTP storage service requests. These requests are transparently
transformed onto IB traffic without modifying workload implementation or
network configuration. This is achieved by 'cloning' a global HTTP service
(BPService) onto the HPC cluster compute host(s) where the workload executes at
scale. Provided the global persistent storage is on the IB network the
HTTP service requests are returned to the workload via IB and not TCP/IP.

The feasibility of this pattern is demonstrated by implementing a prototype
BPService for S3 and a Hybrid-BPService (service metadata is accessed over TCP,
bulk data over IB) for a software repository service. We offer this pattern in
addition to many alternative approaches to address the performance challenges
of executing workloads written to cloud-native design patterns in HPC
environments. Finally, while the initial results are encuraging signficant
validation of any service must be performed under both read and write scenarios
before deployment into production is planned.

\end{abstract}


\section{Introduction} 

Cloud-native software platforms are typically defined by containers,
microservices, infrastructure-as-code, S3 storage, and DevOps practices. Inter-
and intra-platform communication is typically over HTTP over the TCP/IP (IP)
networking standard. HTTP is attractive because it is easy to route, inspect,
proxy, and has a mature readily available ecosystem of tooling.

\begin{figure}[!t]
\centering
\scalebox{0.43}{\import{./figures/}{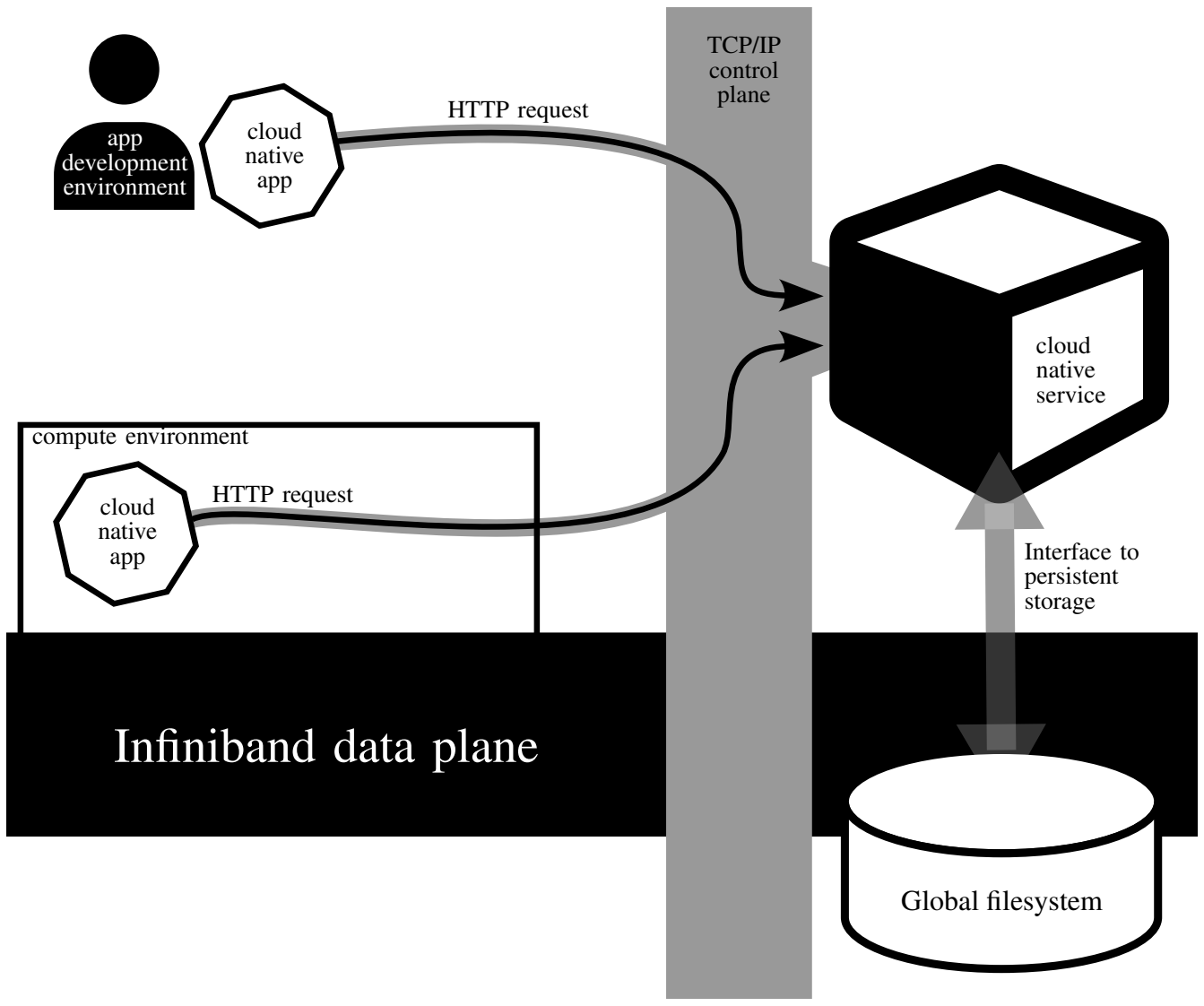}}

\caption{An illustration of the tendency for HTTP calls to seek out TCP
networks. An application is developed following cloud-native software
development patterns. These patterns emphasize HTTP calls for storage and
interprocess communication. When the app is deployed into the compute
environment the HTTP calls remain. The host in the compute environment is
typically configured to send these requests over the available TCP network. In
this configuration the InfiniBand data plane is underutilized.}

\label{fig:problem}
\end{figure}

However, large-scale HPC compute systems are frequently designed for workloads
with coherent memory access and choose remote direct memory access (RDMA)
networks between compute and storage. RDMA deployments commonly use the
InfiniBand (IB) standard, chosen for its high-bandwidth and low-latency. Both
IP and IB can coexist, but communication protocols tend to have a strong
affinity for one or the other. HTTP, in particular, is tightly associated with
IP so cloud-native traffic rarely benefits from an IB fabric.
Figure~\ref{fig:problem} illustrates this behavior.

\section{The Bipartible Pattern}

The Bipartible pattern applies in situations where an environment has been
specified and built for traditional HPC workloads and an optimized experience
for cloud-native type workloads is desired. For example, a container runtime
will typically use HTTP to pull a container into cache before executing. A
workload may request data from an S3 endpoint. With the pattern HTTP calls to
internal cloud-native services are transparently redirected over IB while
keeping HTTP calls to external services are unmodified.

The following conditions must be met for the Bipartible pattern to be viable:
\begin{itemize}

    \item The cloud-native service (BPService) must be hosted within the
		control boundary of the facility.

    \item BPService must be configured to use persistent storage
	and the persistent storage must be available on the InfiniBand data plane.

    \item All nodes in the cluster must be connected to the InfiniBand data plane.

    \item BPService can execute on the host where the workload is running.

\end{itemize}

A BPService is initially deployed in the cluster as a single TCP/HTTP endpoint.
For example, a S3 storage service. Provided the service can be configured to
use persistent storage available on InfiniBand data plane (IBStorage) any
service of interest can become a BPService. If BPService has some fraction of
storage that is not on IBStorage (i.e.  metadata is stored in a database) the
Hybrid-BPService is appropriate. If BPService does not store any significant
data on IBStorage then this service is not a good candidate for this pattern.

\begin{figure}[!t]
\centering
\scalebox{0.43}{\import{./figures/}{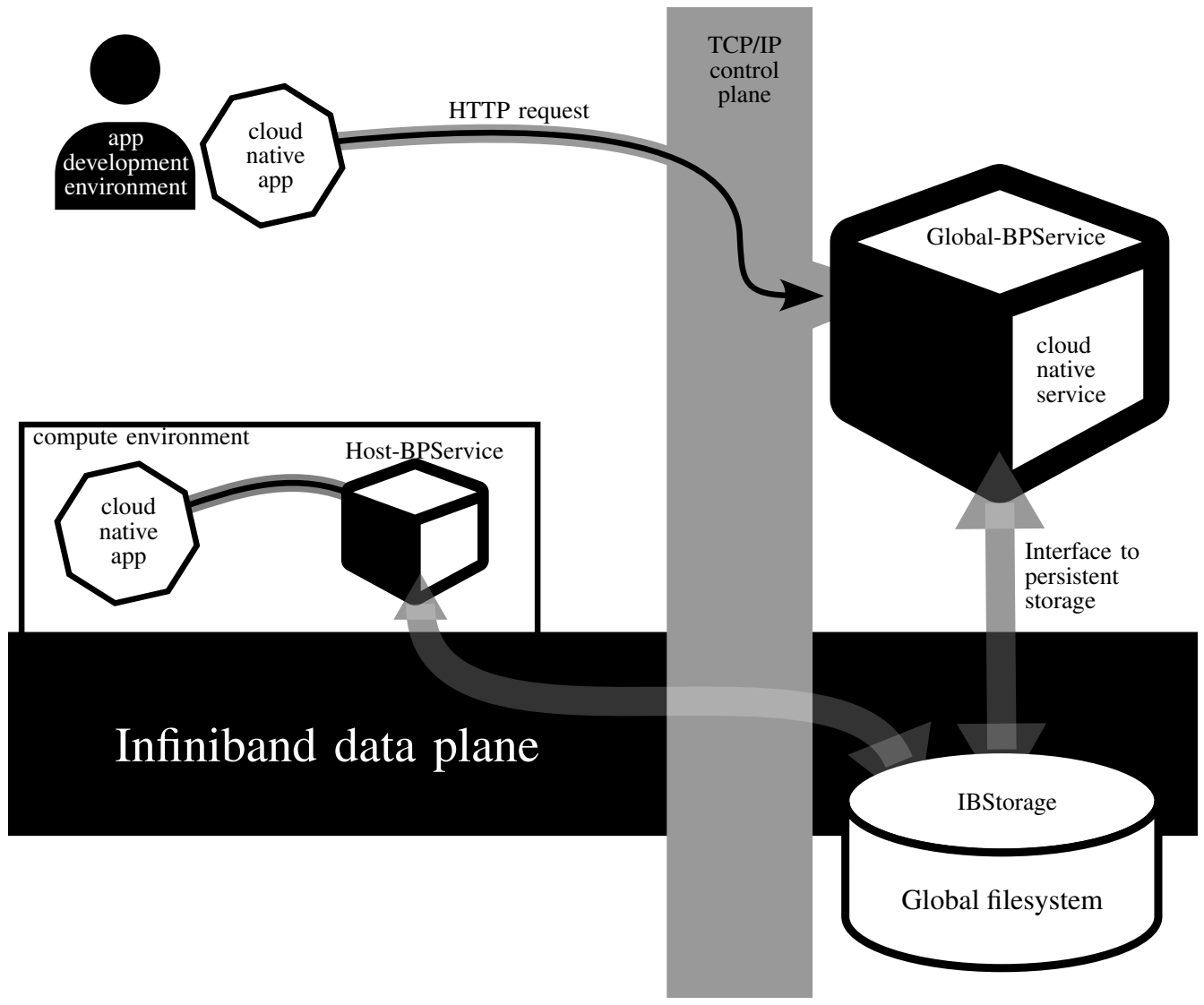}}

\caption{An illustration of the Bipartible architecture pattern. A cloud-
native application is scheduled into a compute environment. HTTP calls are
intercepted on the compute host(s) and redirected to a local clone of the
cloud-native service running on the same host. The internal cloud-native
service (Host-BPService) is configured to use the optimal path from the host to
IBStorage. The suboptimial path to IBStorage via the TCP/IP control plane (and
the Global-BPService) is avoided.}

\label{fig:bipartible}
\end{figure}

BPService has a global (Global-BPService) and host (Host-BPService) instance.
Global-BPService identifies the globally visible instance of the BPService.
Typically, this service will be required to develop the cloud-native app.
Host-BPService is a clone of Global-BPService that is only available from the
compute environment. Global IBStorage access means that Host-BPService is
indistinguishable from Global-BPService and has identical behavior. The host is
configured to intercept HTTP requests (that target Global-BPService) and
redirect them to Host-BPService. Figure~\ref{fig:bipartible} illustrates this
configuration.

Methods to redirect HTTP requests to Host-BPService include but are not limited to:

\begin{itemize}
    \item eBPF-based interception
    \item iptables rules
    \item DNS resolution
    \item Environment variables
    \item Proxy configuration
    \item Network routing
    \item LD\textunderscore PRELOAD/system-call interception
\end{itemize}

The choice of method maybe biased by a number of considerations. These include but are not limited to:

\begin{itemize}
    \item Certificate management
    \item Existing host network configuration
    \item Cluster management configuration tooling
\end{itemize}

\section{Hybrid-Bipartible Pattern}

\begin{figure}[!t]
\centering
\scalebox{0.43}{\import{./figures/}{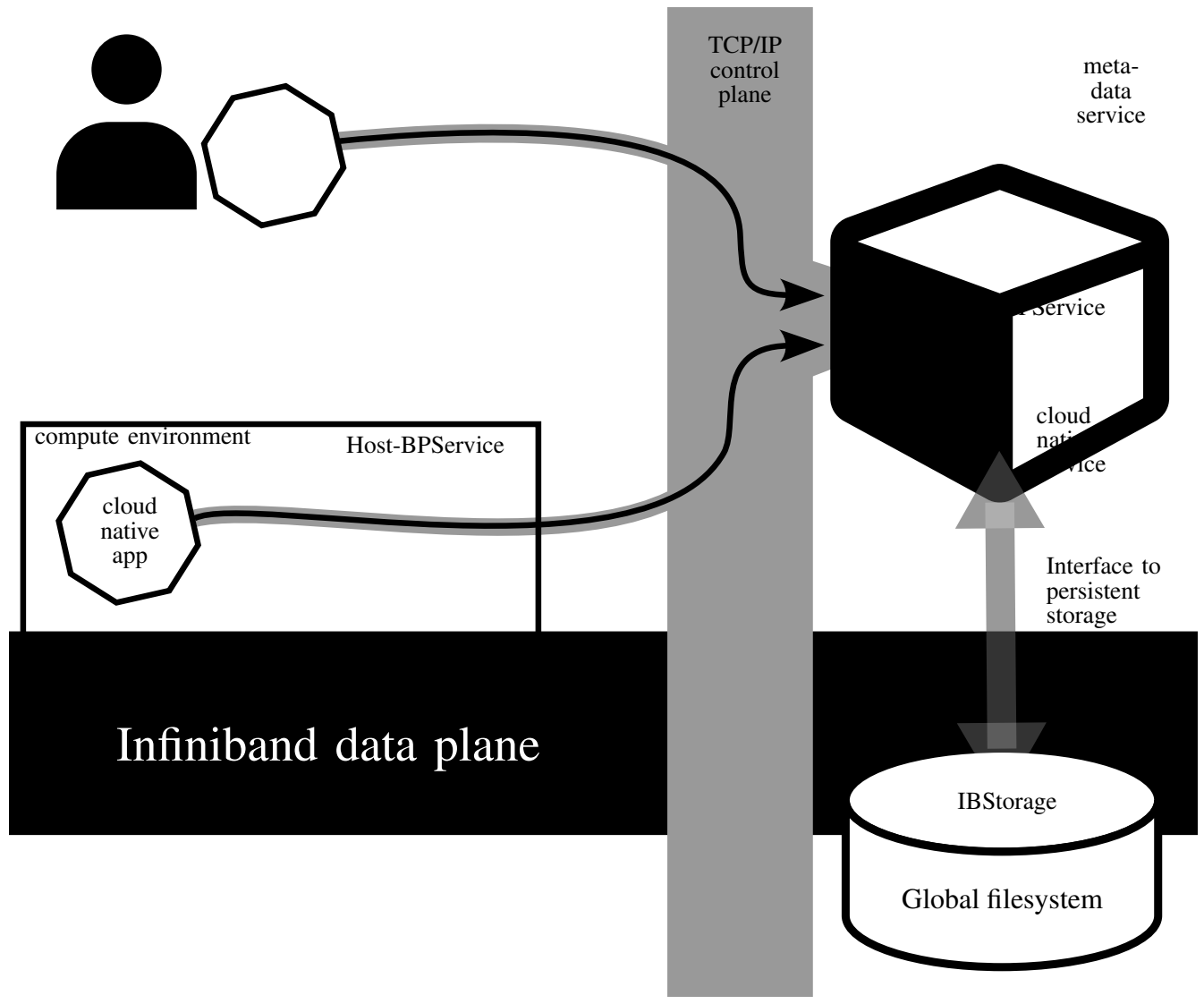}}

\caption{An illustration of the Hybrid-Bipartible architecture pattern. 
This is similar to the Bipartible pattern but includes a metadata component of
BPService that is accessed over TCP/IP. The metadata is accessed over TCP/IP
for both the Global-BPService and the Host-BPService. Bulk data is accessed
over IB for both the Global-BPService and the Host-BPService.}

\label{fig:hybrid-bipartible}
\end{figure}

BPService may use database or key-value storage in addition to IBStorage. The
Hybrid-Bipartible pattern is identical to the standard pattern with the
modification that lightweight non-IBStorage is accessed over TCP/IP. IBStorage for
'bulk' data is still retrieved over IB. A package registry such as Pulp, which
uses SQL for metadata and POSIX file storage for bulk data, is an example of a
service that benefits from this variant. Hybrid-Bipartible is illustrated in 
Figure~\ref{fig:hybrid-bipartible}.

\section{Results}

We evaluated the Bipartible pattern against two candidate services. The current
results are proof-of-concept exercises and are not performance benchmarks or
indications of production readiness. Both evaluations showed encouraging
results in read-only operation. We use an entry in \texttt{/etc/hosts} on the
compute node to redirect HTTP requests to Host-BPService.

\subsection{S3 Object Storage with Garage}

Garage is an open-source, distributed object storage service that can be
self-hosted. We configured it as a single-node Bipartible service. Basic S3
read and write operations succeeded, demonstrating that the Bipartible
redirection mechanism can serve object data from a Lustre global filesystem
over InfiniBand. Attempting to crate a bucket from Host-BPService failed with
the error: 

\begin{lstlisting}
Error: ListBuckets returned InternalError (500): Internal error: Layout not ready.
\end{lstlisting}

Bucket creation may not be a common (or user permitted) activity and the
failure of this function may not prohibit useful deployment of Garage as a
BPService.

\subsection{Container Registry and Software Distribution with Pulp }

Pulp is a container image registry and software distribution system that stores
metadata in a database and bulk data in a Lustre filesystem. Because this metadata
is not stored on a Lustre global filesystem, Pulp requires the Hybrid-Bipartible
pattern, where bulk data is accessed over InfiniBand while metadata remains on
the IP network.

When configured with the Hybrid-Bipartible pattern the Pulp configuration did
not experience any apprently limitations.

\begin{lstlisting}[caption={Pulp Global-BPService container configuration example}, label={lst:global_bpserver}]
IBSTORAGE_PATH="/mnt/path1/bipartiblefs" # IBStorage
DB_HOST="postgres.datacenter.edu"
IMAGE="docker.io/pulp/pulp:latest"

podman run -d \
  --network host \
  -v "${IBSTORAGE_PATH}":/var/lib/pulp:Z \
  "${IMAGE}"
\end{lstlisting}

\begin{lstlisting}[caption={Pulp Host-BPService container configuration example}, label={lst:host_bpserver_pulp}]
# Configuration example for Pulp as Host-BPService
#
# This assumes Global-BPService is already running and has performed a database 
# migration phase.
#
# The following must be shared between Global-BPService and Host-BPService:
# PULP_DB_ENCRYPTION_KEY 
# PULP_SECRET_KEY 

IBSTORAGE_PATH="/mnt/bipartiblefs"  # IBStorage 
DB_HOST="postgres.example.edu"
IMAGE="docker.io/pulp/pulp:latest"

set -- \
  --network host \
  -v "${HOST_DATA_DIR}":/var/lib/pulp:Z \
  -e "PULP_DATABASES__default__HOST=${DB_HOST}" \

podman run -d --name pulp-host-bpservice "$@" "${IMAGE}" pulp-api
podman run -d --name pulp-host-bpservice-content "$@" "${IMAGE}" pulp-content
podman run -d --name pulp-host-bpservice-worker "$@" "${IMAGE}" pulp-worker
\end{lstlisting}

\begin{lstlisting}[caption={Host-BPService \texttt{/etc/hosts} file}, label={lst:host_bpserver_hosts}]
# Configuration example of /etc/hosts for Host-BPService
#
127.0.0.1 localhost

# redirect requests to pulp.example.edu to the service
# running on a container on the local host
127.0.0.1 pulp.example.edu

\end{lstlisting}

\section{Conclusion}

The Bipartible architecture pattern lets cloud-native workloads retain their
HTTP-by-default communication design while exploiting the bandwidth and latency
advantages of InfiniBand. By co-locating a thin service on each host and
relying on global storage on a high-performance fabric, data-heavy traffic is
redirected onto the network fast-path transparently. The Hybrid-Bipartible
variant extends the pattern to services that also need IP-accessible metadata.
Our early results show that the Bipartible pattern is feasible for S3-style
services with POSIX bulk storage. In addition we show Hybrid-Bipartible variant
is the appropriate path for services with a non-POSIX metadata, and demonstrate
this using the Pulp software repository.

Cloud native software not only accesses data over HTTP but often builds on
container images. The Extreme-scale Scientific Software Stack (E4S) container
image typically ranges from 13 GB to over 40 GB. By demonstrating a
container registry (Pulp) as a BPSerivce the authors show the Bipartipble
architecture pattern has the potential to improve container-based workload
startup times without complex network configuration.

\section{Acknowledgements}

An adaptive LLM harness was used to assiet with spelling and minor grammar
critique. The authors would like to thank Eric Roman for valuable feedback on
the development of this concept.


\begin{thebibliography}{1}
\bibitem{E4S}
E4S, ``Ecosystem for HPC and AI Software,'' \url{https://e4s.io/}, accessed 2026.
\bibitem{Lustre}
Lustre, ``Lustre File System,'' \url{https://www.lustre.org/}, accessed 2026.
\bibitem{Garage}
Garage, ``An open-source distributed object storage service,'' \url{https://garagehq.deuxfleurs.fr/}, accessed 2026.
\bibitem{Pulp}
Pulp Project, ``Pulp: Software Repository Management,'' \url{https://pulpproject.org/}, accessed 2026.
\end{thebibliography}
\end{document}